\documentclass[aps,preprint]{revtex4}%
\usepackage{amsfonts}
\usepackage{amsmath}
\usepackage{amssymb}
\usepackage{subfigure}
\usepackage{graphicx}%
\usepackage{color}
\usepackage{array}

\begin{document}
\title{Joule-Thomson expansion of Lower-dimensional black hole}
\author{Jing Liang$^{a,b}$}
\email{jingliang@stu.scu.edu.cn}

\author{Benrong Mu$^{a,b}$}
\email{benrongmu@cdutcm.edu.cn}

\author{Peng Wang$^{b}$}
\email{pengw@scu.edu.cn}

\affiliation{$^{a}$ Physics Teaching and Research section, College of Medical Technology,
Chengdu University of Traditional Chinese Medicine, Chengdu, 611137,
PR China}
\affiliation{$^{b}$Center for Theoretical Physics, College of Physics, Sichuan University, Chengdu, 610064, PR China}

\begin{abstract}
The Joule-Thomson expansion is extended to the lower-dimensional regime by considering the rotating BTZ metric in the (2+1)-dimensional space-time. Specifically, the properties of three important aspects of the Joule-Thomson expansion, namely the Joule-Thomson coefficient, the inversion curve and the isenthalpic curve are focused on.
The divergence point of the Joule-Thomson coefficient and the zero point of the Hawking temperature are studied. The inversion temperature curves and isenthalpic curves in the $T-P$ plane are obtained and the cooling-heating regions are determined. Furthermore, the minimum inversion temperature is found to be zero, and the black hole becomes an extremal black hole. The ratio between the minimum inversion temperature and the critical temperature for the BTZ black hole doesn't exist, since the BTZ black hole does not have the critical behavior in $P_c$, $T_c$ and $V_c$.

\end{abstract}

\maketitle
\tableofcontents{}

\bigskip{}



\section{Introduction}

Since Hawking's discovery that black holes can thermodynamically emit particles, black holes have become a topic that attracted widespread attention. Various thermodynamic properties of black holes have been extensively studied until now \cite{intro-Bekenstein:1972tm,intro-Bekenstein:1973ur,intro-Bardeen:1973gs,intro-Bekenstein:1974ax,intro-Hawking:1974rv,intro-Hawking:1974sw}. Notably, it has been found that black holes as thermodynamic systems have many similarities with universal thermodynamic systems. For black holes in AdS space, these similarities are more obvious and precise. The study of AdS black hole thermodynamics started with the seminal paper by Hawking and Page, in which a phase transition was found between a Schwarzschild-AdS black hole and a thermal AdS space \cite{intro-Hawking:1982dh}. Subsequently, attention was focused on Hawking radiation and phase transition in various black holes \cite{intro-Jiang:2005ba,intro-Boos:2019vcz,intro-Jiang:2005xb,intro-Setare:2006hq,intro-Jiang:2007pn,intro-Myung:2006sq,intro-Jiang:2007wj,intro-Banerjee:2010qk}.

In recent years, thermodynamic studies of black holes in AdS space have focused on extended phase space. In extended phase space, the cosmological constant and its conjugate quantities are considered as thermodynamic pressure and volume, respectively
\begin{equation}
P=-\frac{\Lambda}{8\pi}=\frac{\left(D-1\right)\left(D-2\right)}{16\text{\ensuremath{\pi}}l^{2}},V=\left(\frac{\partial M}{\partial P}\right)_{S,Q},
\label{eqn:PV1}
\end{equation}
where $l$ is the AdS space radius and the black hole mass $M$ is treated as the enthalpy \cite{intro-Kubiznak:2016qmn}. Subsequently, many papers have explored various thermodynamic aspects of black holes in extended phase space, such as the phase transition \cite{intro-Altamirano:2013uqa,intro-Wei:2014hba}, the critical phenomenon \cite{intro-Wei:2012ui,intro-Banerjee:2011cz,intro-Niu:2011tb}, compressibility \cite{intro-Dolan:2011jm,intro-Dolan:2013dga}, heat engine efficiency \cite{intro-Johnson:2014yja,intro-Hendi:2017bys}, and the weak cosmic censorship conjecture \cite{intro-Gwak:2017kkt,intro-Gwak:2018akg,intro-Yang:2020czk,intro-Chen:2019nsr,intro-Wang:2019jzz,intro-Mu:2019bim,intro-Liang:2021voh,intro-Chen:2020zps,intro-Mu:2020szg,intro-Liang:2020hjz,intro-Chen:2019pdj,intro-Gwak:2019asi,intro-Liang:2020uul,intro-Zeng:2019jta,intro-Wang:2019dzl,intro-Han:2019kjr,intro-Han:2019lfs,intro-Zeng:2019aao}.

Hawking and Page studied the phase transition between the  Schwarzschild-AdS black hole and thermal AdS space \cite{intro-Hawking:1982dh}. It was shown that black holes in AdS space have common properties with the general thermodynamic system. This relationship is further enhanced in the extended phase space \cite{intro-Kubiznak:2012wp}. The Joule-Thomson expansion of the black hole was first studied in the literature \cite{intro-Okcu:2016tgt}. Subsequently, the Joule-Thomson expansion was further extended to other kinds of black holes, such as $d$-dimensional charged AdS black holes \cite{intro-Mo:2018rgq}, Kerr-AdS black holes \cite{intro-Okcu:2017qgo}, regular(Bardeen)-AdS black holes \cite{intro-Pu:2019bxf}, RN-AdS black holes in $f(R)$ gravity \cite{intro-Chabab:2018zix}, quintessence RN-AdS black holes \cite{intro-Ghaffarnejad:2018exz}, Bardeen-AdS black holes \cite{intro-Li:2019jcd} and other black holes \cite{intro-Rizwan:2018mpy,intro-Yekta:2019wmt,intro-Cisterna:2018jqg,intro-Lan:2018nnp,intro-Cao:2021dcq,intro-Meng:2020csd,intro-Bi:2020vcg,intro-Guo:2020zcr,intro-Rostami:2019ivr,intro-Haldar:2018cks,intro-Mo:2018qkt,intro-Nam:2019zyk,intro-Ghaffarnejad:2018tpr,intro-Kuang:2018goo,intro-Zhao:2018kpz,intro-Lan:2019kak,intro-Guo:2019pzq,intro-Ranjbari:2019ktp,intro-Sadeghi:2020bon,intro-Farsam:2020pfl}. In these papers, the inversion curves, the isenthalpic curves and the heating-cooling regions in the $T-P$ plane for different black holes were given. The results of these papers show that the inversion curves are similar for different black hole systems.
Until now, all the work has focused on the space-times with dimension $D \geq 4$, while the fact that $D < 4$ dimensions remains to be explored. However, the case of $D < 4$ is well worth studying. The renewed interest in low-dimensional gravity theory in the last decade or so has been inspired by the confluence of evidence suggesting an effective two-dimensional Planck regime. Recently, a black hole model inspired by the generalized uncertainty principle (GUP) was shown to exhibit reduced dimensional properties within the sub-Planckian masses limit \cite{intro-Carr:2015nqa}. This suggesting that the physics of quantum black holes is actually low-dimensional (a similarly emergent two-dimensional space-time was mentioned in Ref. \cite{intro-Nicolini:2012fy}). Since then, the expansion of the thermodynamic phase space of low-dimensional black holes have attracted a lot of attention.
In this paper, we extend the present study of Joule-Thomson expansion to the rotating BTZ black hole.
The BTZ black hole is a solution of the Einstein field equation in the (2+1)-dimensional space-time and describes a rotating AdS geometry \cite{intro-Banados:1992wn,intro-Banados:1992gq}. In previous work, many papers have investigated the thermodynamic properties of BTZ black holes, such as gravitational perturbation \cite{intro-Rocha:2011wp}, stability of the event horizon \cite{intro-Wang:1996hp}, the entropy \cite{intro-Carlip:1998qw}, space-time duality \cite{intro-Ho:1997st}, statistical entropy \cite{intro-Saida:1999ec,intro-Park:2006zw} and others \cite{intro-Mann:2008rx,intro-Chen:2018yah,intro-Zeng:2019huf,intro-Duztas:2016xfg,intro-Mu:2019jjw,intro-Zhao:2006bc,intro-Akbar:2007zz}.

The paper is organized as follows. In Section \ref{sec:T}, we review the thermodynamics of the rotating BTZ black hole. Then in Section \ref{sec:JT}, we discuss the Joule-Thomson expansion of a rotating BTZ black hole, including the Joule-Thomson coefficient, the inversion curves and the isenthalpic curves. Our results are discussed in Section \ref{sec:con}.

\section{The rotating BTZ black hole}
\label{sec:T}
The BTZ black hole is a solution of Einstein field equation in (2+1)-dimensional space-time with a negative cosmological constant $\Lambda=-l^{2}$, and $l$ is AdS radius. The corresponding action with a negative term $\Lambda$ is expressed as
\begin{equation}
S=\frac{1}{2\pi}\int\sqrt{-g}\left[R+2\Lambda\right].
\label{eqn:action}
\end{equation}
The BTZ black hole solution to the action $\left(\ref{eqn:action}\right)$ is given by \cite{intro-Banados:1992wn,intro-Banados:1992gq}
\begin{equation}
ds^{2}=-f(r)dt^{2}+\frac{1}{f\left(r\right)}dr^{2}+r^{2}\left(d\varphi+N_{\phi}\left(r\right)dt\right)^{2},
\end{equation}
where
\begin{equation}
f(r)=-2m+\frac{r^{2}}{l^{2}}+\frac{J^{2}}{4r^{2}},
\end{equation}
\begin{equation}
N_{\phi}\left(r\right)=-\frac{J}{2r^{2}}.
\end{equation}
When the black hole is  nonextremal, the equation $f(r)= 0$ has two positive real roots $r_+$ and $r_-$, where the largest root $r_+$ denotes the radius of the event horizon.
When the black hole is extremal, $f(r)= 0$ has only one root $r_+$.
The metric parameters $m$ is related to the black hole mass $M$, and can be expressed in
terms of $r_+$ as
\begin{equation}
M=\frac{m}{4}=\frac{J^{2}}{32r_+^{2}}+\frac{r_+^{2}}{8l^{2}}.
\label{eqn:M1}
\end{equation}
The mass of the black hole in terms of entropy $S$, angular momentum $J$ and pressure $P$ is given by
\begin{equation}
M=\frac{\pi^{2}J^{2}}{128S^{2}}+\frac{4PS^{2}}{\pi}.
\label{eqn:M2}
\end{equation}
The corresponding thermodynamic quantities are then \cite{M-Chen:2012mh,M-Cai:1996df,M-Cai:1999dz,M-Cai:1998ep,M-Zou:2014gla}
\begin{equation}
S=\frac{\pi r_{+}}{2},T=\frac{r_{+}}{2\pi l^{2}}-\frac{J^{2}}{8\pi r_{+}^{3}},
\label{eqn:ST}
\end{equation}
\begin{equation}
P=\frac{1}{8\pi l^{2}},V=\pi r^{2},\Omega=\frac{J}{16r_{+}^{2}}.
\label{eqn:PV}
\end{equation}
A black hole as a stable thermodynamic system can be discussed in two phase spaces. In normal phase space, the cosmological constant is considered as constant, the state parameters of the black hole satisfy the first law of thermodynamics
\begin{equation}
dM=TdS+\Omega dJ,
\label{eqn:dMn}
\end{equation}
However, contrary to the usual first law of thermodynamics, the VdP term is missing in Eq. $\left(\ref{eqn:dMn}\right)$.  Inspired by this, the cosmological constant is considered as the pressure of the black hole, and the relationship between them is shown in Eq. $\left(\ref{eqn:PV1}\right)$. Therefore, in the extended phase space, the first law of thermodynamics is \cite{M-Wang:2006eb}
\begin{equation}
dM=TdS+VdP+\Omega dJ,
\label{eqn:dM}
\end{equation}
and the Smarr formula is
\begin{equation}
0=TS-2PV+\Omega J.
\label{eqn:smarr}
\end{equation}
The mass of the black hole $M$ is defined as its enthalpy
\begin{equation}
M=U+PV.
\end{equation}
\section{Joule-Thomson expansion}
\label{sec:JT}
In Joule-Thomson expansion, the gas is passed at high pressure through a porous plug or small valve in a low-pressure section of an adiabatic tube, and the enthalpy remains constant during the expansion. Expansion is characterized by a change in temperature relative to pressure. The Joule-Thomson coefficient $\mu$, which characterizes the expansion process, is given by
\begin{equation}
\mu=\left(\frac{\partial T}{\partial P}\right)_{H}.
\end{equation}
From Eqs.  $\left(\ref{eqn:ST}\right)$, $\left(\ref{eqn:PV}\right)$, $\left(\ref{eqn:dM}\right)$ and $\left(\ref{eqn:smarr}\right)$, the heat capacity at constant pressure is
\begin{equation}
C_{P}=T\left(\frac{\partial S}{\partial T}\right)_{P,J}=\frac{1}{2}\pi r_{+}\left(1-\frac{4J^{2}}{3J^{2}+32\pi Pr_{+}^{4}}\right),
\end{equation}
and one can obtain
\begin{equation}
\mu=\left(\frac{\partial T}{\partial P}\right)_{H}=\frac{1}{C_{P}}\left[T\left(\frac{\partial V}{\partial T}\right)_{P}-V\right]=\frac{2r\left(5J^{2}-32\pi Pr_{+}^{4}\right)}{J^{2}-32\pi Pr_{+}^{4}}.
\end{equation}

\begin{figure}
  \centering
  \subfigure[{Joule-Thomson coefficient versus the event horizon.}]{
  \includegraphics[width=0.45\textwidth]{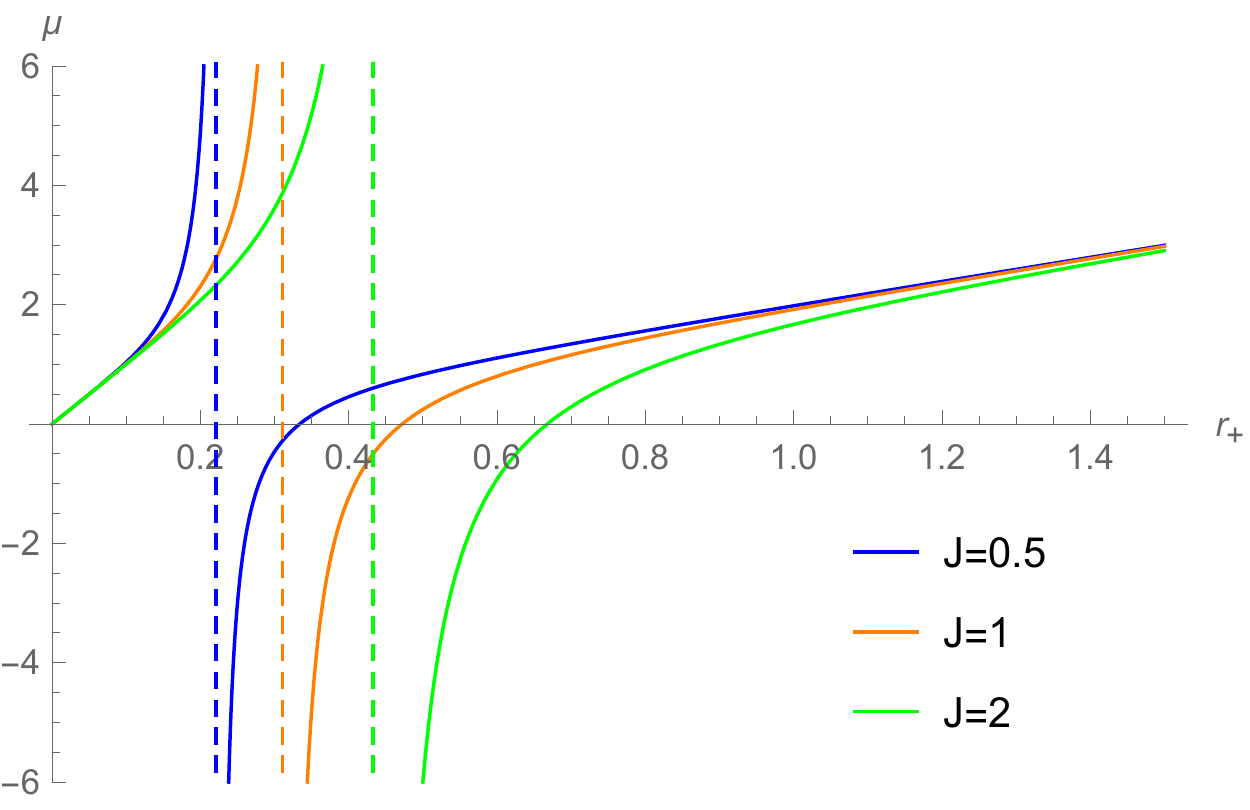}\label{fig:con}}
  \subfigure[{Hawking temperature $T$ versus the event horizon.}]{
  \includegraphics[width=0.45\textwidth]{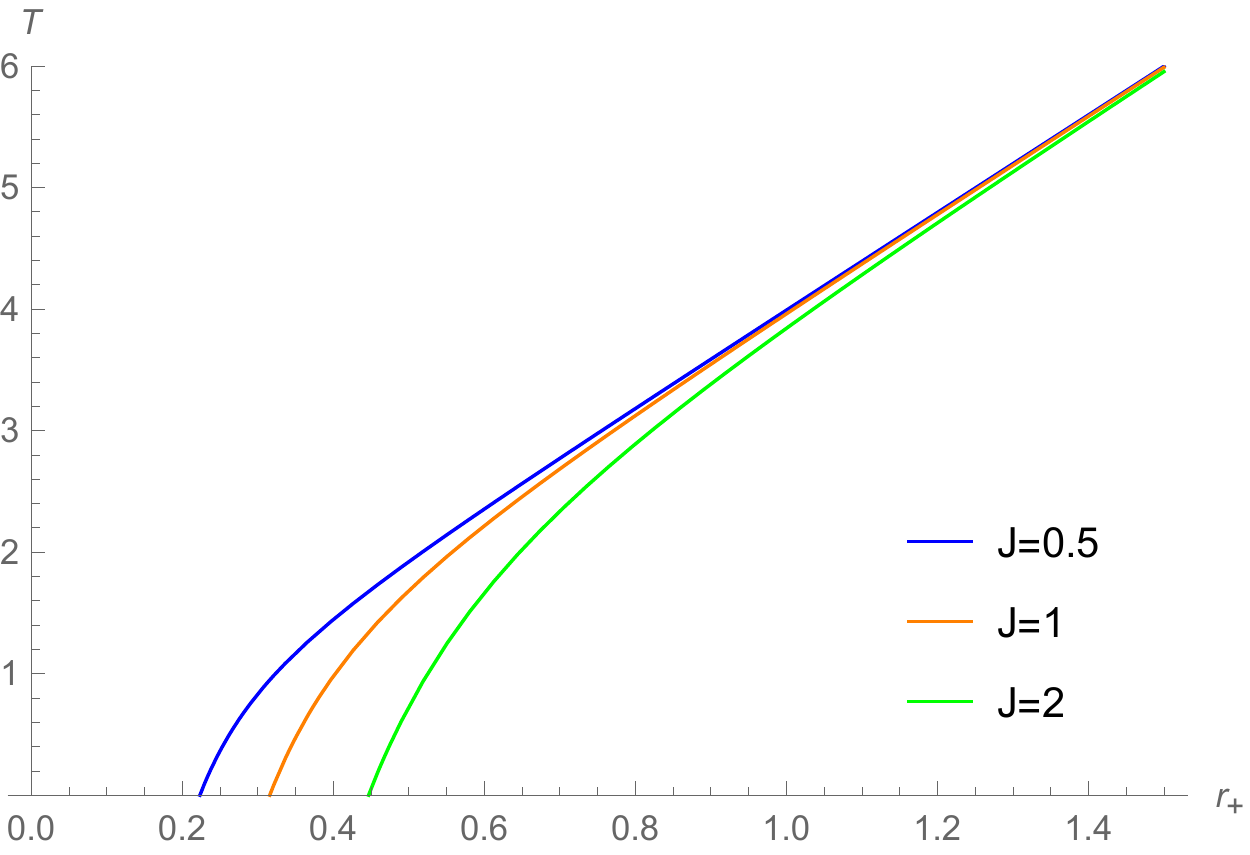}\label{fig:Tr}}\\
  \caption{Joule-Thomson coefficient and Hawking temperature $T$ versus the event horizon. Here, $P=1$.}\label{conT}
\end{figure}

In Fig. \ref{conT}, the Joule-Thomson coefficient and Hawking temperature versus the horizon are shown. We fix the pressure $P=1$ and the angular momentum $J$ as 0.5, 1, 2. There exist both a divergence point and a zero point for different $J$. By comparing these two figures, it is easy to see that the divergence point of the Joule-Thomson coefficient is consistent with the zero point of Hawking temperature. The divergence point here reveals the Hawking temperature and corresponds to extremal black holes.

\begin{figure}[htb]
\centering
\includegraphics[scale=0.7]{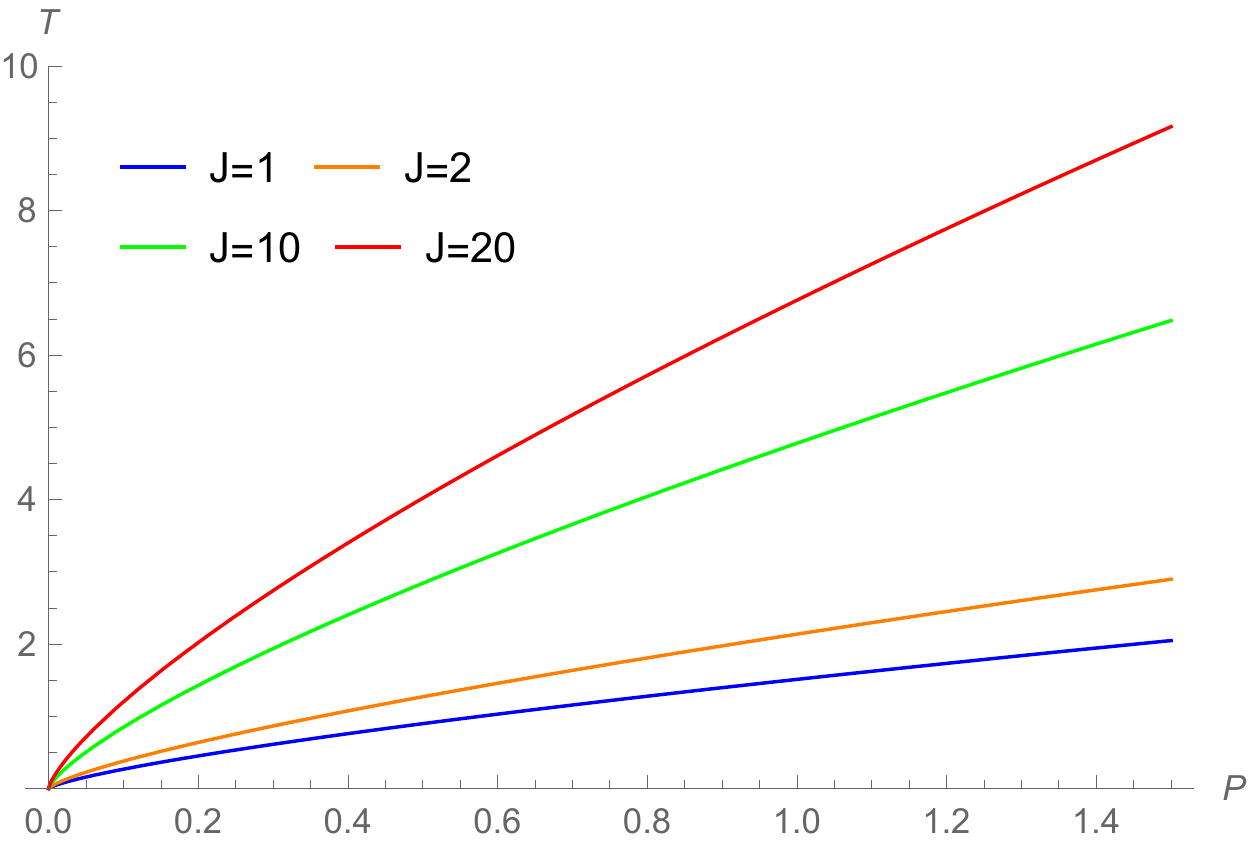}
\caption{Inversion curves for BTZ black hole.}
\label{fig:TP}
\end{figure}

\begin{figure}[htb]
\begin{center}
\subfigure[{$J=1$, $M=1.5,2.0,2.5,3.0$.}]{
\includegraphics[width=0.45\textwidth]{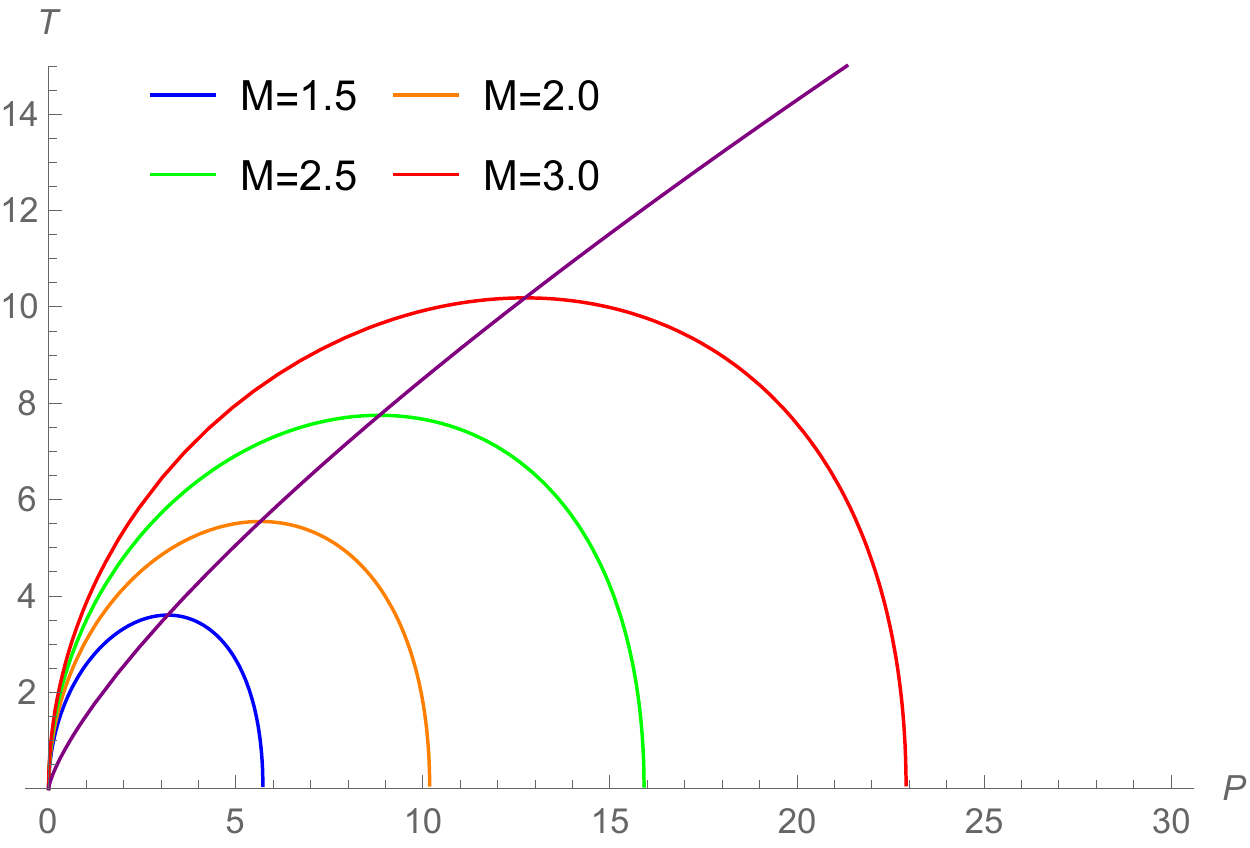}\label{fig:H1}}
\subfigure[{$J=2$, $M=2.5,3.0,3.5,4.0$.}]{
\includegraphics[width=0.45\textwidth]{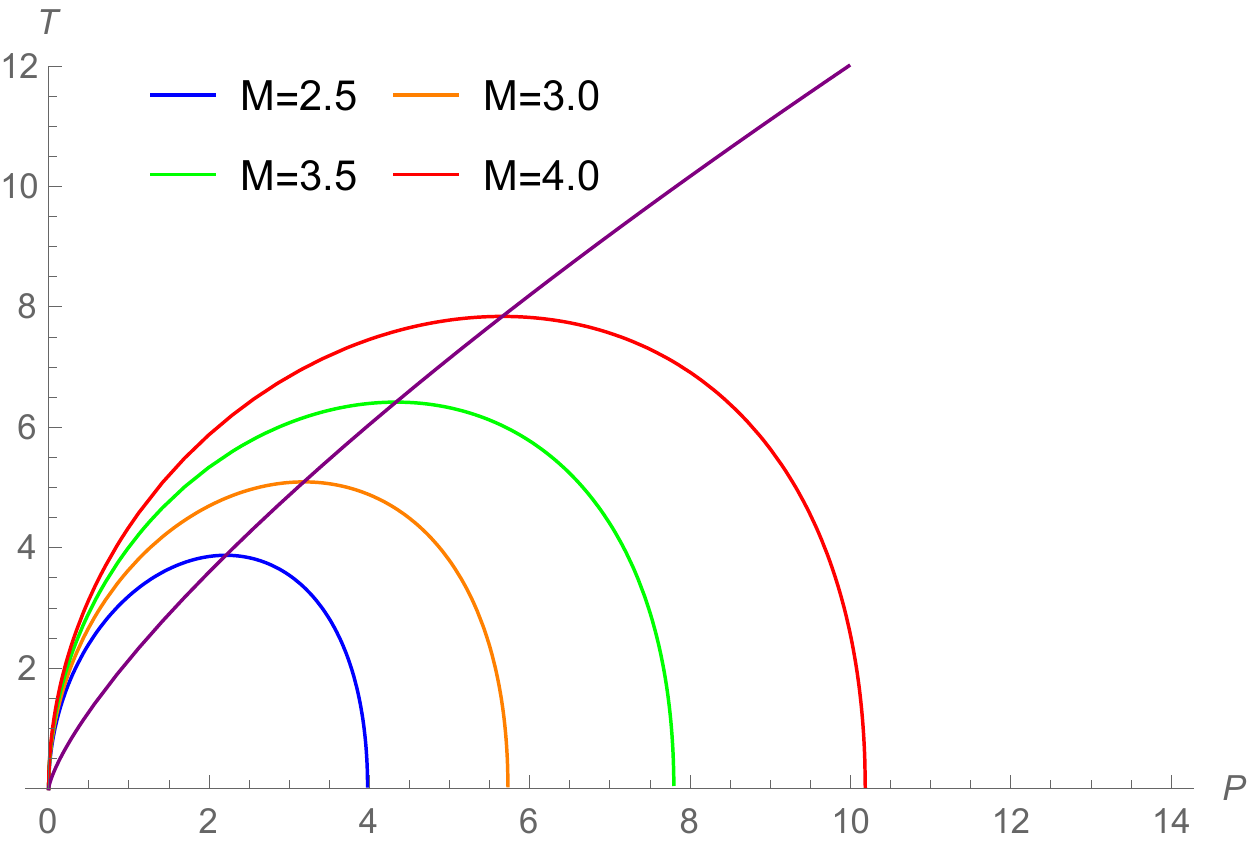}\label{fig:H2}}
\subfigure[{$J=10$, $M=10.5,11.0,11.5,12.0$.}]{
\includegraphics[width=0.45\textwidth]{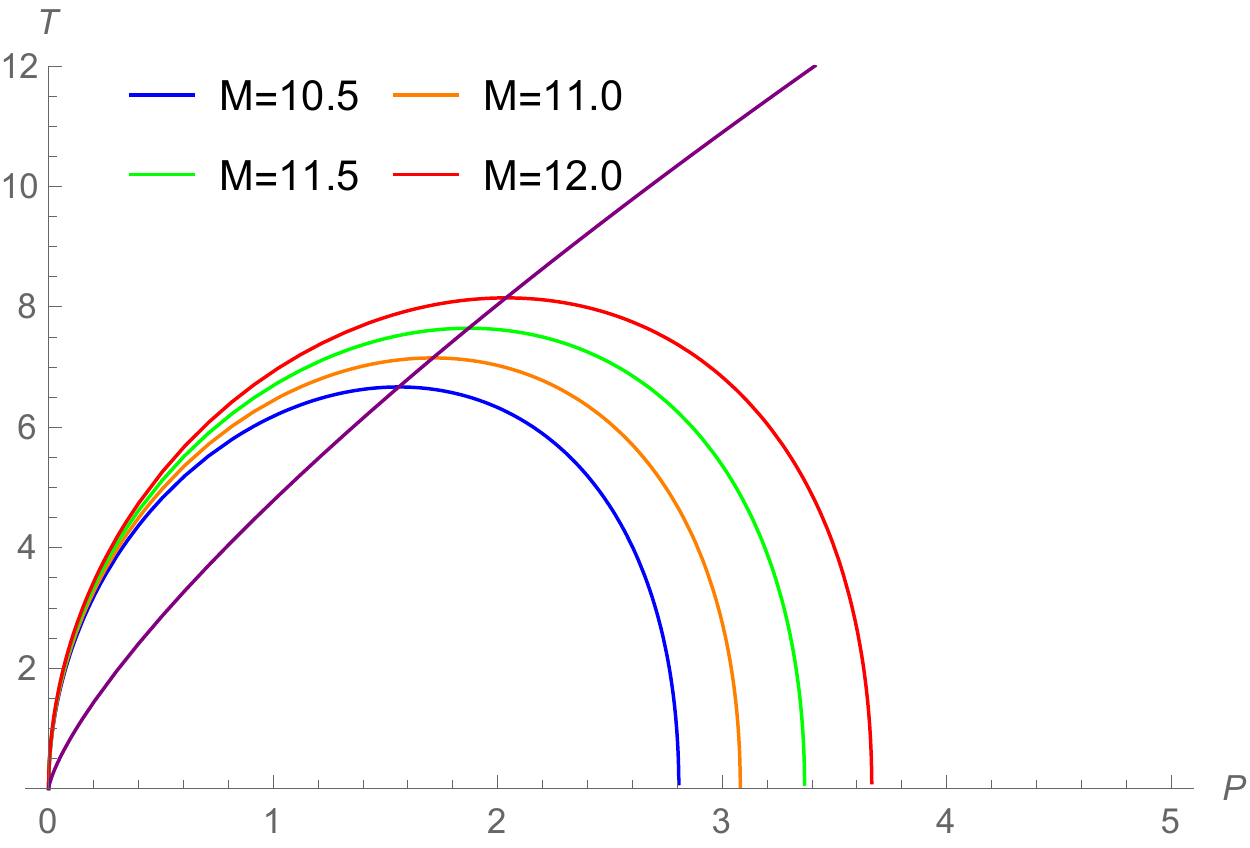}\label{fig:H10}}
\subfigure[{$J=20$, $M=20.5,21.0,21.5,22.0$.}]{
\includegraphics[width=0.45\textwidth]{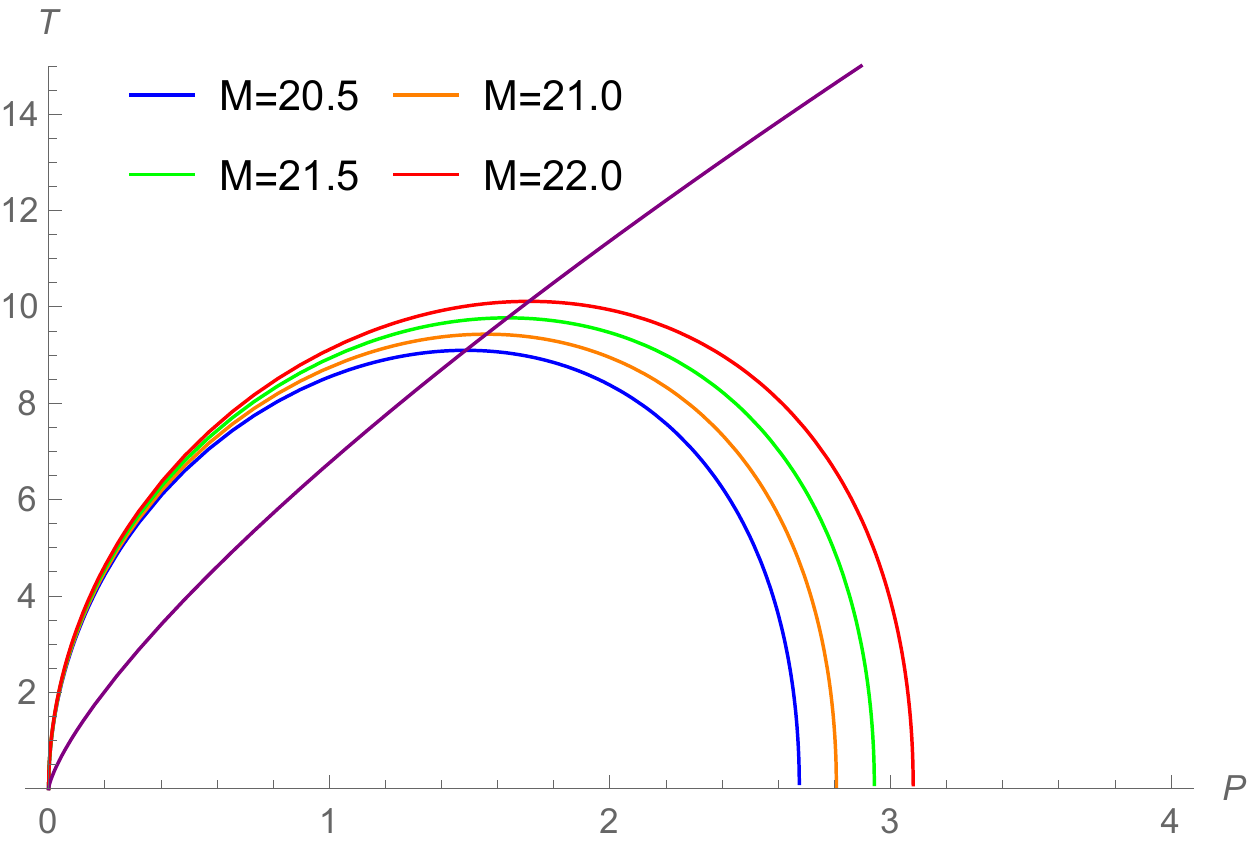}\label{fig:H20}}
\end{center}
\caption{The inversion and isenthalpic curves for the BTZ black hole. From bottom to top, the value of $M$ corresponding to the isenthalpic curve increases. The purple lines are the inversion curves.}%
\label{fig:JTH}
\end{figure}
The black hole equation of state is given by
\begin{equation}
T=4Pr_+-\frac{J^{2}}{8\pi r_{+}^{3}}.
\label{eqn:T1}
\end{equation}
Then, the inversion temperature takes on the form as
\begin{equation}
T_{i}=V\left(\frac{\partial V}{\partial T}\right)_{P}=\frac{3J^{2}}{16\pi r_{+}^{3}}+2Pr_{+}.
\label{eqn:TI1}
\end{equation}
Subtracting Eq. $\left(\ref{eqn:TI1}\right)$ from $\left(\ref{eqn:T1}\right)$ yields
\begin{equation}
\frac{5J^{2}}{16\pi r_{+}^{3}}-2P_{i}r_{+}=0.
\end{equation}
Solving this equation for $r_+$ yields four roots, only one of which is physically meaningful, the other roots are complex or negative. The positive and real root is
\begin{equation}
r_{+}=\left(\frac{5J^{2}}{32\pi P_{i}}\right)^{\frac{1}{4}}.
\end{equation}
Substitute this root into Eq. $\left(\ref{eqn:T1}\right)$ at $P=P_i$, the inversion temperature is given by
\begin{equation}
T_{i}=\frac{4\left(\frac{2}{5}\right)^{3/4}\sqrt{J}P_{i}^{3/4}}{\sqrt[4]{\pi}}.
\end{equation}
By setting $P_i=0$ in the above equation, the minimum of the inversion temperature is
\begin{equation}
T_{i}^{min}=0,
\end{equation}
which means the black hole becomes an extremal black hole. As shown in Ref. \cite{JT-Sadeghi:2015out}, the BTZ black hole does not have $P-V$ critical behavior in the critical pressure $P_c$, the critical temperature $T_c$ and the critical volume $V_c$. Thus, the BTZ hole is always thermodynamically stable, and the ratio between minimum inversion temperature and critical temperature doesn't exist.

From Fig. \ref{fig:TP}, the inversion temperature increases monotonically with increasing inversion pressure, but the slope of the inversion curve decreases with increasing inversion pressure. In addition, in contrast to the Van der Waals fluids, there is only a lower inversion curve and this curve does not terminate at any point.
Joule-Thomson expansion occurs in an isenthalpic process. For a black hole, the enthalpy is the mass $M$. The isenthalpic curve can be obtained by Eqs. $\left(\ref{eqn:M1}\right)$ and $\left(\ref{eqn:PV}\right)$. In Fig. \ref{fig:JTH}, both the inversion curve and the isenthalpic curve for the black hole are shown. Above the inversion curves, the slope of the isenthalpic curve is positive, so that cooling occurs above the inversion curve. Below the inversion curve, the sign of the slope of the isenthalpic curve changes, and heating occurs in this region.

\section{Conclusion}
\label{sec:con}
The Joule-Thomson expansion of the BTZ black hole in the extended phase space was investigated in this paper, where the cosmological constant is considered as pressure and the black hole mass is treated as enthalpy. In Fig. \ref{conT}, Joule-Thomson coefficient $\mu$ and Hawking temperature $T$ versus the event horizon are presented. The divergence point of the Joule-Thomson coefficient is consistent with the zero point of the Hawking temperature, which corresponds to the extremal black hole. In Fig. \ref{fig:TP}, the inversion curves for BTZ black hole are shown. The isenthalpic curves and inversion curves are given in Fig. \ref{fig:JTH}. The results show that the inversion curve always has a positive slope, which means that BTZ black hole always cools above the inversion curve during the expansion. The inversion curve can be used to distinguish different values of the cooling and heating regions. Moreover, the minimum of the inversion temperature is 0, and the black hole becomes an extremal black hole. As shown in Ref. \cite{JT-Sadeghi:2015out}, it is well known that the BTZ black hole does not have the critical behavior in $P_c$, $T_c$ and $V_c$. Therefore, the ratio between minimum inversion temperature and critical temperature doesn't exist. In Table \ref{tab:ratio}, the existence of critical behavior and the ratio of the minimum inversion temperature to the critical temperature of the Van der Waals fluid, the charged AdS black hole, the d-dimensional AdS black hole, the Gauss-Bonnet black hole, the torus-like black hole and the BTZ black hole are listed.

\begin{table}[htb]
\begin{centering}
\begin{tabular}{|m{1.6in}<{\centering}|m{0.8in}<{\centering}|m{0.9in}<{\centering}|m{0.7in}<{\centering}|}
\hline
type  &the critical behavior & ratio & literature\tabularnewline
\hline
Van der Waals fluid & exist & 0.75 & \cite{intro-Okcu:2016tgt} \tabularnewline
RN-AdS BH & exist & 0.5 & \cite{intro-Ghaffarnejad:2018exz} \tabularnewline
d-dimensional AdS BH  & exist & $<$0.5 & \cite{intro-Mo:2018rgq} \tabularnewline
Gauss-Bonnet BH & exist & 0.4765 & \cite{intro-Lan:2018nnp} \tabularnewline
torus-like BH     & not exist  & not exist & \cite{con-Liang:2021elg} \tabularnewline
BTZ BH & not exist & not exist & \tabularnewline
\hline
\end{tabular}
\par\end{centering}
\caption{{\footnotesize{}{}{}{}The existence of critical behavior and the ratio of the minimum inversion temperature to the critical temperature of various black holes.}}
\label{tab:ratio}
\end{table}

In this paper, we have focused only on the Joule-Thomson expansion of BTZ black holes in low-dimensional black holes. These results are related to many other interesting problems that deserve further investigation. In the near future, it is reasonable to study the extended phase space thermodynamics of low-dimensional black holes. This motivates us to further discuss the effect of dimensionality on the thermodynamic properties of black holes. We hope to see more studies on the Joule-Thomson expansion of low-dimensional black holes in the future.

\begin{acknowledgments}
We are grateful to Wei Hong, Haitang Yang, Jun Tao, Deyou Chen and Xiaobo Guo for useful discussions. This work is supported in part by NSFC (Grant No. 11747171), Natural Science Foundation of Chengdu University of TCM (Grants nos. ZRYY1729 and ZRYY1921), Discipline Talent Promotion Program of /Xinglin Scholars(Grant no.QNXZ2018050) and the key fund project for Education Department of Sichuan (Grantno. 18ZA0173).
\end{acknowledgments}


\begin{thebibliography}{}
\bibitem{intro-Bekenstein:1972tm}
J.~D.~Bekenstein,
``Black holes and the second law,''
Lett. Nuovo Cim. \textbf{4}, 737-740 (1972)
doi:10.1007/BF02757029

\bibitem{intro-Bekenstein:1973ur}
J.~D.~Bekenstein,
``Black holes and entropy,''
Phys. Rev. D \textbf{7}, 2333-2346 (1973)
doi:10.1103/PhysRevD.7.2333

\bibitem{intro-Bardeen:1973gs}
J.~M.~Bardeen, B.~Carter and S.~W.~Hawking,
``The Four laws of black hole mechanics,''
Commun. Math. Phys. \textbf{31}, 161-170 (1973)
doi:10.1007/BF01645742

\bibitem{intro-Bekenstein:1974ax} J.~D.~Bekenstein,
``Generalized second law of thermodynamics in black hole physics,''
Phys. Rev. D \textbf{9}, 3292-3300 (1974)
doi:10.1103/PhysRevD.9.3292

\bibitem{intro-Hawking:1974rv}
S.~W.~Hawking,
``Black hole explosions,''
Nature \textbf{248}, 30-31 (1974)
doi:10.1038/248030a0

\bibitem{intro-Hawking:1974sw}
S.~W.~Hawking,
``Particle Creation by Black Holes,''
Commun. Math. Phys. \textbf{43}, 199-220 (1975)
[erratum: Commun. Math. Phys. \textbf{46}, 206 (1976)]
doi:10.1007/BF02345020

\bibitem{intro-Hawking:1982dh}
S.~W.~Hawking and D.~N.~Page,
``Thermodynamics of Black Holes in anti-De Sitter Space,''
Commun. Math. Phys. \textbf{87}, 577 (1983)
doi:10.1007/BF01208266




\bibitem{intro-Jiang:2005ba}
Q.~Q.~Jiang, S.~Q.~Wu and X.~Cai,
``Hawking radiation as tunneling from the Kerr and Kerr-Newman black holes,''
Phys. Rev. D \textbf{73}, 064003 (2006)
[erratum: Phys. Rev. D \textbf{73}, 069902 (2006)]
doi:10.1103/PhysRevD.73.064003
[arXiv:hep-th/0512351 [hep-th]].

\bibitem{intro-Boos:2019vcz}
J.~Boos, V.~P.~Frolov and A.~Zelnikov,
``Ghost-free modification of the Polyakov action and Hawking radiation,''
Phys. Rev. D \textbf{100}, no.10, 104008 (2019)
doi:10.1103/PhysRevD.100.104008
[arXiv:1909.01494 [hep-th]].

\bibitem{intro-Jiang:2005xb}
Q.~Q.~Jiang and S.~Q.~Wu,
``Hawking radiation of charged particles as tunneling from Reissner-Nordstr\"om-de Sitter black holes with a global monopole,''
Phys. Lett. B \textbf{635}, 151-155 (2006)
[erratum: Phys. Lett. B \textbf{639}, 684-684 (2006)]
doi:10.1016/j.physletb.2006.06.009
[arXiv:hep-th/0511123 [hep-th]].

\bibitem{intro-Setare:2006hq}
M.~R.~Setare,
``Gauge and gravitational anomalies and Hawking radiation of rotating BTZ black holes,''
Eur. Phys. J. C \textbf{49}, 865-868 (2007)
doi:10.1140/epjc/s10052-006-0148-8
[arXiv:hep-th/0608080 [hep-th]].

\bibitem{intro-Jiang:2007pn}
Q.~Q.~Jiang, S.~Q.~Wu and X.~Cai,
``Hawking radiation from the (2+1)-dimensional BTZ black holes,''
Phys. Lett. B \textbf{651}, 58-64 (2007)
doi:10.1016/j.physletb.2007.05.058
[arXiv:hep-th/0701048 [hep-th]].

\bibitem{intro-Myung:2006sq}
Y.~S.~Myung,
``Phase transition between the BTZ black hole and AdS space,''
Phys. Lett. B \textbf{638}, 515-518 (2006)
doi:10.1016/j.physletb.2006.04.024
[arXiv:gr-qc/0603051 [gr-qc]].

\bibitem{intro-Jiang:2007wj}
Q.~Q.~Jiang, S.~Q.~Wu and X.~Cai,
``Hawking radiation from the dilatonic black holes via anomalies,''
Phys. Rev. D \textbf{75}, 064029 (2007)
[erratum: Phys. Rev. D \textbf{76}, 029904 (2007)]
doi:10.1103/PhysRevD.76.029904
[arXiv:hep-th/0701235 [hep-th]].

\bibitem{intro-Banerjee:2010qk}
R.~Banerjee, S.~K.~Modak and S.~Samanta,
``Glassy Phase Transition and Stability in Black Holes,''
Eur. Phys. J. C \textbf{70}, 317-328 (2010)
doi:10.1140/epjc/s10052-010-1443-y
[arXiv:1002.0466 [hep-th]].

\bibitem{intro-Kubiznak:2016qmn}
D.~Kubiznak, R.~B.~Mann and M.~Teo,
``Black hole chemistry: thermodynamics with Lambda,''
Class. Quant. Grav. \textbf{34}, no.6, 063001 (2017)
doi:10.1088/1361-6382/aa5c69
[arXiv:1608.06147 [hep-th]].

\bibitem{intro-Altamirano:2013uqa}
N.~Altamirano, D.~Kubiz\v{n}\'ak, R.~B.~Mann and Z.~Sherkatghanad,
``Kerr-AdS analogue of triple point and solid/liquid/gas phase transition,''
Class. Quant. Grav. \textbf{31}, 042001 (2014)
doi:10.1088/0264-9381/31/4/042001
[arXiv:1308.2672 [hep-th]].

\bibitem{intro-Wei:2014hba}
S.~W.~Wei and Y.~X.~Liu,
``Triple points and phase diagrams in the extended phase space of charged Gauss-Bonnet black holes in AdS space,''
Phys. Rev. D \textbf{90}, no.4, 044057 (2014)
doi:10.1103/PhysRevD.90.044057
[arXiv:1402.2837 [hep-th]].

\bibitem{intro-Wei:2012ui}
S.~W.~Wei and Y.~X.~Liu,
``Critical phenomena and thermodynamic geometry of charged Gauss-Bonnet AdS black holes,''
Phys. Rev. D \textbf{87}, no.4, 044014 (2013)
doi:10.1103/PhysRevD.87.044014
[arXiv:1209.1707 [gr-qc]].

\bibitem{intro-Banerjee:2011cz}
R.~Banerjee and D.~Roychowdhury,
``Critical phenomena in Born-Infeld AdS black holes,''
Phys. Rev. D \textbf{85}, 044040 (2012)
doi:10.1103/PhysRevD.85.044040
[arXiv:1111.0147 [gr-qc]].

\bibitem{intro-Niu:2011tb}
C.~Niu, Y.~Tian and X.~N.~Wu,
``Critical Phenomena and Thermodynamic Geometry of RN-AdS Black Holes,''
Phys. Rev. D \textbf{85}, 024017 (2012)
doi:10.1103/PhysRevD.85.024017
[arXiv:1104.3066 [hep-th]].

\bibitem{intro-Dolan:2011jm}
B.~P.~Dolan,
``Compressibility of rotating black holes,''
Phys. Rev. D \textbf{84}, 127503 (2011)
doi:10.1103/PhysRevD.84.127503
[arXiv:1109.0198 [gr-qc]].

\bibitem{intro-Dolan:2013dga}
B.~P.~Dolan,
``The compressibility of rotating black holes in D-dimensions,''
Class. Quant. Grav. \textbf{31}, 035022 (2014)
doi:10.1088/0264-9381/31/3/035022
[arXiv:1308.5403 [gr-qc]].

\bibitem{intro-Johnson:2014yja}
C.~V.~Johnson,
``Holographic Heat Engines,''
Class. Quant. Grav. \textbf{31}, 205002 (2014)
doi:10.1088/0264-9381/31/20/205002
[arXiv:1404.5982 [hep-th]].

\bibitem{intro-Hendi:2017bys}
S.~H.~Hendi, B.~Eslam Panah, S.~Panahiyan, H.~Liu and X.~H.~Meng,
``Black holes in massive gravity as heat engines,''
Phys. Lett. B \textbf{781}, 40-47 (2018)
doi:10.1016/j.physletb.2018.03.072
[arXiv:1707.02231 [hep-th]].

\bibitem{intro-Gwak:2017kkt}
B.~Gwak,
``Thermodynamics with Pressure and Volume under Charged Particle Absorption,''
JHEP \textbf{11}, 129 (2017)
doi:10.1007/JHEP11(2017)129
[arXiv:1709.08665 [gr-qc]].

\bibitem{intro-Gwak:2018akg}
B.~Gwak,
``Weak Cosmic Censorship Conjecture in Kerr-(Anti-)de Sitter Black Hole with Scalar Field,''
JHEP \textbf{09}, 081 (2018)
doi:10.1007/JHEP09(2018)081
[arXiv:1807.10630 [gr-qc]].

\bibitem{intro-Yang:2020czk}
S.~J.~Yang, J.~J.~Wan, J.~Chen, J.~Yang and Y.~Q.~Wang,
``Weak cosmic censorship conjecture for the novel $4D$ charged Einstein-Gauss-Bonnet black hole with test scalar field and particle,''
Eur. Phys. J. C \textbf{80}, no.10, 937 (2020)
doi:10.1140/epjc/s10052-020-08511-9
[arXiv:2004.07934 [gr-qc]].


\bibitem{intro-Chen:2019nsr}
D.~Chen, W.~Yang and X.~Zeng,
``Thermodynamics and weak cosmic censorship conjecture in Reissner-Nordstr\"om anti-de Sitter black holes with scalar field,''
Nucl. Phys. B \textbf{946}, 114722 (2019)
doi:10.1016/j.nuclphysb.2019.114722
[arXiv:1901.05140 [hep-th]].

\bibitem{intro-Wang:2019jzz}
P.~Wang, H.~Wu and H.~Yang,
``Thermodynamics of nonlinear electrodynamics black holes and the validity of weak cosmic censorship at charged particle absorption,''
Eur. Phys. J. C \textbf{79}, no.7, 572 (2019)
doi:10.1140/epjc/s10052-019-7090-z


\bibitem{intro-Mu:2019bim}
B.~Mu and J.~Tao,
``Minimal Length Effect on Thermodynamics and Weak Cosmic Censorship Conjecture in anti-de Sitter Black Holes via Charged Particle Absorption,''
Advances in High Energy Physics,
vol. 2020, Article ID 2612946, 9 pages, 2020.
doi:10.1155/2020/2612946.
[arXiv:1906.10544 [gr-qc]]


\bibitem{intro-Liang:2021voh}
J.~Liang, X.~Guo and B.~Mu,
``Thermodynamics with pressure and volume of black holes based on two assumptions under scalar field scattering,''
[arXiv:2101.11414 [gr-qc]].

\bibitem{intro-Chen:2020zps}
D.~Chen and S.~Zeng,
``Overcharging problem and thermodynamics in extended phase spaces,''
Nucl. Phys. B \textbf{957}, 115089 (2020)
doi:10.1016/j.nuclphysb.2020.115089
[arXiv:2003.08102 [gr-qc]].




\bibitem{intro-Mu:2020szg}
B.~Mu, J.~Liang and X.~Guo,
``Thermodynamics with pressure and volume of 4D Gauss-Bonnet AdS Black Holes under the scalar field,''
[arXiv:2011.00273 [gr-qc]].

\bibitem{intro-Liang:2020hjz}
J.~Liang, X.~Guo, D.~Chen and B.~Mu,
``Remarks on the weak cosmic censorship conjecture of RN-AdS black holes with cloud of strings and quintessence under the scalar field,''
doi:10.1016/j.nuclphysb.2021.115335
[arXiv:2008.08327 [gr-qc]].

\bibitem{intro-Chen:2019pdj}
D.~Chen,
``Thermodynamics and weak cosmic censorship conjecture in extended phase spaces of anti-de Sitter black holes with particles\textquoteright{} absorption,''
Eur. Phys. J. C \textbf{79}, no.4, 353 (2019)
doi:10.1140/epjc/s10052-019-6874-5
[arXiv:1902.06489 [hep-th]].

\bibitem{intro-Gwak:2019asi}
B.~Gwak,
``Weak Cosmic Censorship with Pressure and Volume in Charged Anti-de Sitter Black Hole under Charged Scalar Field,''
JCAP \textbf{08}, 016 (2019)
doi:10.1088/1475-7516/2019/08/016
[arXiv:1901.05589 [gr-qc]].


\bibitem{intro-Liang:2020uul}
J.~Liang, B.~Mu and J.~Tao,
``Thermodynamics and overcharging problem in the extended phase spaces of charged AdS black holes with cloud of strings and quintessence under charged particle absorption,''
Chin. Phys. C \textbf{45}, no.2, 023121 (2021)
doi:10.1088/1674-1137/abd085
[arXiv:2008.09512 [gr-qc]].

\bibitem{intro-Zeng:2019jta}
X.~X.~Zeng and H.~Q.~Zhang,
``Thermodynamics and weak cosmic censorship conjecture in Born-Infeld-anti-de Sitter black holes,''
Chin. Phys. C \textbf{45}, no.2, 025112 (2021)
doi:10.1088/1674-1137/abd088
[arXiv:1901.04247 [hep-th]].

\bibitem{intro-Wang:2019dzl}
P.~Wang, H.~Wu and H.~Yang,
``Thermodynamics and Weak Cosmic Censorship Conjecture in Nonlinear Electrodynamics Black Holes via Charged Particle Absorption,''
[arXiv:1904.12365 [gr-qc]].

\bibitem{intro-Han:2019kjr}
Y.~W.~Han, X.~X.~Zeng and Y.~Hong,
``Thermodynamics and weak cosmic censorship conjecture of the torus-like black hole,''
Eur. Phys. J. C \textbf{79}, no.3, 252 (2019)
doi:10.1140/epjc/s10052-019-6771-y
[arXiv:1901.10660 [hep-th]].


\bibitem{intro-Han:2019lfs}
Y.~W.~Han, M.~J.~Lan and X.~X.~Zeng,
``Thermodynamics and weak cosmic censorship conjecture in (2+1)-dimensional regular black hole with nonlinear electrodynamics sources,''
Eur. Phys. J. Plus \textbf{135}, no.2, 172 (2020)
doi:10.1140/epjp/s13360-020-00186-1
[arXiv:1903.03764 [gr-qc]].

\bibitem{intro-Zeng:2019aao}
X.~X.~Zeng and H.~Q.~Zhang,
``Thermodynamics and weak cosmic censorship conjecture in the Kerr-AdS black hole,''
Nucl. Phys. B \textbf{959}, 115162 (2020)
doi:10.1016/j.nuclphysb.2020.115162
[arXiv:1905.01618 [gr-qc]].



\bibitem{intro-Kubiznak:2012wp}
D.~Kubiznak and R.~B.~Mann,
``P-V criticality of charged AdS black holes,''
JHEP \textbf{07}, 033 (2012)
doi:10.1007/JHEP07(2012)033
[arXiv:1205.0559 [hep-th]].

\bibitem{intro-Okcu:2016tgt}
\"O.~\"Okc\"u and E.~Ayd\i{}ner,
``Joule\textendash{}Thomson expansion of the charged AdS black holes,''
Eur. Phys. J. C \textbf{77}, no.1, 24 (2017)
doi:10.1140/epjc/s10052-017-4598-y
[arXiv:1611.06327 [gr-qc]].

\bibitem{intro-Mo:2018rgq}
J.~X.~Mo, G.~Q.~Li, S.~Q.~Lan and X.~B.~Xu,
``Joule-Thomson expansion of $d$-dimensional charged AdS black holes,''
Phys. Rev. D \textbf{98}, no.12, 124032 (2018)
doi:10.1103/PhysRevD.98.124032
[arXiv:1804.02650 [gr-qc]].

\bibitem{intro-Okcu:2017qgo}
\"O.~\"Okc\"u and E.~Ayd\i{}ner,
``Joule\textendash{}Thomson expansion of Kerr\textendash{}AdS black holes,''
Eur. Phys. J. C \textbf{78}, no.2, 123 (2018)
doi:10.1140/epjc/s10052-018-5602-x
[arXiv:1709.06426 [gr-qc]].

\bibitem{intro-Pu:2019bxf}
J.~Pu, S.~Guo, Q.~Q.~Jiang and X.~T.~Zu,
``Joule-Thomson expansion of the regular(Bardeen)-AdS black hole,''
Chin. Phys. C \textbf{44}, no.3, 035102 (2020)
doi:10.1088/1674-1137/44/3/035102
[arXiv:1905.02318 [gr-qc]].

\bibitem{intro-Chabab:2018zix}
M.~Chabab, H.~El Moumni, S.~Iraoui, K.~Masmar and S.~Zhizeh,
``Joule-Thomson Expansion of RN-AdS Black Holes in $f(R)$ gravity,''
LHEP \textbf{02}, 05 (2018)
doi:10.31526/LHEP.2.2018.02
[arXiv:1804.10042 [gr-qc]].

\bibitem{intro-Ghaffarnejad:2018exz}
H.~Ghaffarnejad, E.~Yaraie and M.~Farsam,
``Quintessence Reissner Nordstr\"om Anti de Sitter Black Holes and Joule Thomson effect,''
Int. J. Theor. Phys. \textbf{57}, no.6, 1671-1682 (2018)
doi:10.1007/s10773-018-3693-7
[arXiv:1802.08749 [gr-qc]].


\bibitem{intro-Li:2019jcd}
C.~Li, P.~He, P.~Li and J.~B.~Deng,
``Joule-Thomson expansion of the Bardeen-AdS black holes,''
Gen. Rel. Grav. \textbf{52}, no.5, 50 (2020)
doi:10.1007/s10714-020-02704-z
[arXiv:1904.09548 [gr-qc]].



\bibitem{intro-Rizwan:2018mpy}
A.~Rizwan C.L., N.~Kumara A., D.~Vaid and K.~M.~Ajith,
``Joule-Thomson expansion in AdS black hole with a global monopole,''
Int. J. Mod. Phys. A \textbf{33}, no.35, 1850210 (2019)
doi:10.1142/S0217751X1850210X
[arXiv:1805.11053 [gr-qc]].


\bibitem{intro-Yekta:2019wmt}
D.~Mahdavian Yekta, A.~Hadikhani and \"O.~\"Okc\"u,
``Joule-Thomson expansion of charged AdS black holes in Rainbow gravity,''
Phys. Lett. B \textbf{795}, 521-527 (2019)
doi:10.1016/j.physletb.2019.06.049
[arXiv:1905.03057 [hep-th]].


\bibitem{intro-Cisterna:2018jqg}
A.~Cisterna, S.~Q.~Hu and X.~M.~Kuang,
``Joule-Thomson expansion in AdS black holes with momentum relaxation,''
Phys. Lett. B \textbf{797}, 134883 (2019)
doi:10.1016/j.physletb.2019.134883
[arXiv:1808.07392 [gr-qc]].



\bibitem{intro-Lan:2018nnp}
S.~Q.~Lan,
``Joule-Thomson expansion of charged Gauss-Bonnet black holes in AdS space,''
Phys. Rev. D \textbf{98}, no.8, 084014 (2018)
doi:10.1103/PhysRevD.98.084014
[arXiv:1805.05817 [gr-qc]].

\bibitem{intro-Cao:2021dcq}
Y.~Cao, H.~Feng, W.~Hong and J.~Tao,
``Joule-Thomson Expansion of RN-AdS Black Hole Immersed in Perfect Fluid Dark Matter,''
[arXiv:2101.08199 [gr-qc]].

\bibitem{intro-Meng:2020csd}
Y.~Meng, J.~Pu and Q.~Q.~Jiang,
``P-V criticality and Joule-Thomson expansion of charged AdS black holes in the Rastall gravity,''
Chin. Phys. C \textbf{44}, no.6, 065105 (2020)
doi:10.1088/1674-1137/44/6/065105

\bibitem{intro-Bi:2020vcg}
S.~Bi, M.~Du, J.~Tao and F.~Yao,
``Joule-Thomson expansion of Born-Infeld AdS black holes,''
Chin. Phys. C \textbf{45}, no.2, 025109 (2021)
doi:10.1088/1674-1137/abcf23
[arXiv:2006.08920 [gr-qc]].

\bibitem{intro-Guo:2020zcr}
S.~Guo, Y.~Han and G.~P.~Li,
``Thermodynamic of the charged AdS black holes in Rastall gravity: P \ensuremath{-} V critical and Joule\textendash{}Thomson expansion,''
Mod. Phys. Lett. A \textbf{35}, no.14, 2050113 (2020)
doi:10.1142/S0217732320501138


\bibitem{intro-Rostami:2019ivr}
M.~Rostami, J.~Sadeghi, S.~Miraboutalebi, A.~A.~Masoudi and B.~Pourhassan,
``Charged accelerating AdS black hole of $f(R)$ gravity and the Joule\textendash{}Thomson expansion,''
Int. J. Geom. Meth. Mod. Phys. \textbf{17}, no.09, 2050136 (2020)
doi:10.1142/S0219887820501364
[arXiv:1908.08410 [gr-qc]].

\bibitem{intro-Haldar:2018cks}
A.~Haldar and R.~Biswas,
``Joule-Thomson expansion of five-dimensional Einstein-Maxwell-Gauss-Bonnet-AdS black holes,''
EPL \textbf{123}, no.4, 40005 (2018)
doi:10.1209/0295-5075/123/40005


\bibitem{intro-Mo:2018qkt}
J.~X.~Mo and G.~Q.~Li,
``Effects of Lovelock gravity on the Joule\textendash{}Thomson expansion,''
Class. Quant. Grav. \textbf{37}, no.4, 045009 (2020)
doi:10.1088/1361-6382/ab60b9
[arXiv:1805.04327 [gr-qc]].

\bibitem{intro-Nam:2019zyk}
C.~H.~Nam,
``Heat engine efficiency and Joule-Thomson expansion of non-linear charged AdS black hole in massive gravity,''
Gen. Rel. Grav. \textbf{53}, no.3, 30 (2021)
doi:10.1007/s10714-021-02787-2
[arXiv:1906.05557 [gr-qc]].

\bibitem{intro-Ghaffarnejad:2018tpr}
H.~Ghaffarnejad and E.~Yaraie,
``Effects of a cloud of strings on the extended phase space of Einstein\textendash{}Gauss\textendash{}Bonnet AdS black holes,''
Phys. Lett. B \textbf{785}, 105-111 (2018)
doi:10.1016/j.physletb.2018.08.017
[arXiv:1806.06687 [gr-qc]].

\bibitem{intro-Kuang:2018goo}
X.~M.~Kuang, B.~Liu and A.~\"Ovg\"un,
``Nonlinear electrodynamics AdS black hole and related phenomena in the extended thermodynamics,''
Eur. Phys. J. C \textbf{78}, no.10, 840 (2018)
doi:10.1140/epjc/s10052-018-6320-0
[arXiv:1807.10447 [gr-qc]].

\bibitem{intro-Zhao:2018kpz}
Z.~W.~Zhao, Y.~H.~Xiu and N.~Li,
``Throttling process of the Kerr\textendash{}Newman\textendash{}anti-de Sitter black holes in the extended phase space,''
Phys. Rev. D \textbf{98}, no.12, 124003 (2018)
doi:10.1103/PhysRevD.98.124003
[arXiv:1805.04861 [gr-qc]].

\bibitem{intro-Lan:2019kak}
S.~Q.~Lan,
``Joule-Thomson expansion of neutral AdS black holes in massive gravity,''
Nucl. Phys. B \textbf{948}, 114787 (2019)
doi:10.1016/j.nuclphysb.2019.114787

\bibitem{intro-Guo:2019pzq}
S.~Guo, Y.~Han and G.~P.~Li,
``Joule-Thomson expansion of a specific black hole in different dimensions,''
[arXiv:1912.09590 [hep-th]].

\bibitem{intro-Ranjbari:2019ktp}
H.~Ranjbari, M.~Sadeghi, M.~Ghanaatian and G.~Forozani,
``Critical behavior of AdS Gauss\textendash{}Bonnet massive black holes in the presence of external string cloud,''
Eur. Phys. J. C \textbf{80}, no.1, 17 (2020)
doi:10.1140/epjc/s10052-019-7592-8
[arXiv:1911.10803 [hep-th]].

\bibitem{intro-Sadeghi:2020bon}
J.~Sadeghi and R.~Toorandaz,
``Joule-Thomson expansion of hyperscaling violating black holes with spherical and hyperbolic horizons,''
Nucl. Phys. B \textbf{951}, 114902 (2020)
doi:10.1016/j.nuclphysb.2019.114902


\bibitem{intro-Farsam:2020pfl}
M.~Farsam, E.~Yaraie, H.~Ghaffarnejad and E.~Ghasami,
``Cooling-heating phase transition for 4D AdS Bardeen Gauss-Bonnet Black Hole,''
[arXiv:2010.05697 [hep-th]].


\bibitem{intro-Carr:2015nqa}
B.~J.~Carr, J.~Mureika and P.~Nicolini,
``Sub-Planckian black holes and the Generalized Uncertainty Principle,''
JHEP \textbf{07}, 052 (2015)
doi:10.1007/JHEP07(2015)052
[arXiv:1504.07637 [gr-qc]].

\bibitem{intro-Nicolini:2012fy}
P.~Nicolini and E.~Spallucci,
``Holographic screens in ultraviolet self-complete quantum gravity,''
Adv. High Energy Phys. \textbf{2014}, 805684 (2014)
doi:10.1155/2014/805684
[arXiv:1210.0015 [hep-th]].


\bibitem{intro-Banados:1992wn}
M.~Banados, C.~Teitelboim and J.~Zanelli,
``The Black hole in three-dimensional space-time,''
Phys. Rev. Lett. \textbf{69}, 1849-1851 (1992)
doi:10.1103/PhysRevLett.69.1849
[arXiv:hep-th/9204099 [hep-th]].

\bibitem{intro-Banados:1992gq}
M.~Banados, M.~Henneaux, C.~Teitelboim and J.~Zanelli,
``Geometry of the (2+1) black hole,''
Phys. Rev. D \textbf{48}, 1506-1525 (1993)
[erratum: Phys. Rev. D \textbf{88}, 069902 (2013)]
doi:10.1103/PhysRevD.48.1506
[arXiv:gr-qc/9302012 [gr-qc]].





\bibitem{intro-Rocha:2011wp}
J.~V.~Rocha and V.~Cardoso,
``Gravitational perturbation of the BTZ black hole induced by test particles and weak cosmic censorship in AdS spacetime,''
Phys. Rev. D \textbf{83}, 104037 (2011)
doi:10.1103/PhysRevD.83.104037
[arXiv:1102.4352 [gr-qc]].

\bibitem{intro-Wang:1996hp}
B.~Wang, R.~K.~Su and P.~K.~N.~Yu,
``Stability of the event horizon in (2+1)-dimensional black holes,''
Phys. Rev. D \textbf{54}, 7298-7302 (1996)
doi:10.1103/PhysRevD.54.7298

\bibitem{intro-Carlip:1998qw}
S.~Carlip,
``What we don't know about BTZ black hole entropy,''
Class. Quant. Grav. \textbf{15}, 3609-3625 (1998)
doi:10.1088/0264-9381/15/11/020
[arXiv:hep-th/9806026 [hep-th]].

\bibitem{intro-Ho:1997st}
J.~Ho, W.~T.~Kim and Y.~J.~Park,
``Space-time duality of BTZ black hole,''
J. Korean Phys. Soc. \textbf{33}, S541-S544 (1998)
[arXiv:gr-qc/9902047 [gr-qc]].

\bibitem{intro-Saida:1999ec}
H.~Saida and J.~Soda,
``Statistical entropy of BTZ black hole in higher curvature gravity,''
Phys. Lett. B \textbf{471}, 358-366 (2000)
doi:10.1016/S0370-2693(99)01405-7
[arXiv:gr-qc/9909061 [gr-qc]].

\bibitem{intro-Park:2006zw}
M.~I.~Park,
``BTZ Black Hole with Higher Derivatives, the Second Law of Thermodynamics, and Statistical Entropy: A New Proposal,''
Phys. Rev. D \textbf{77}, 126012 (2008)
doi:10.1103/PhysRevD.77.126012
[arXiv:hep-th/0609027 [hep-th]].







\bibitem{intro-Mann:2008rx}
R.~B.~Mann, J.~J.~Oh and M.~I.~Park,
``The Role of Angular Momentum and Cosmic Censorship in the (2+1)-Dimensional Rotating Shell Collapse,''
Phys. Rev. D \textbf{79}, 064005 (2009)
doi:10.1103/PhysRevD.79.064005
[arXiv:0812.2297 [hep-th]].

\bibitem{intro-Chen:2018yah}
D.~Chen,
``Weak cosmic censorship conjecture in BTZ black holes with scalar fields,''
Chin. Phys. C \textbf{44}, no.1, 015101 (2020)
doi:10.1088/1674-1137/44/1/015101
[arXiv:1812.03459 [gr-qc]].

\bibitem{intro-Zeng:2019huf}
X.~X.~Zeng, Y.~W.~Han and D.~Y.~Chen,
``Thermodynamics and weak cosmic censorship conjecture of BTZ black holes in extended phase space,''
Chin. Phys. C \textbf{43}, no.10, 105104 (2019)
doi:10.1088/1674-1137/43/10/105104
[arXiv:1901.08915 [gr-qc]].


\bibitem{intro-Duztas:2016xfg}
K.~D\"uzta\c{s},
``Overspinning BTZ black holes with test particles and fields,''
Phys. Rev. D \textbf{94}, no.12, 124031 (2016)
doi:10.1103/PhysRevD.94.124031
[arXiv:1701.07241 [gr-qc]].

\bibitem{intro-Mu:2019jjw}
B.~Mu, J.~Tao and P.~Wang,
``Free-fall Rainbow BTZ Black Hole,''
Phys. Lett. B \textbf{800}, 135098 (2020)
doi:10.1016/j.physletb.2019.135098
[arXiv:1906.11703 [gr-qc]].

\bibitem{intro-Zhao:2006bc}
R.~Zhao and S.~L.~Zhang,
``Canonical entropy of three-dimensional BTZ black hole,''
Phys. Lett. B \textbf{641}, 318-322 (2006)
doi:10.1016/j.physletb.2006.08.068
[arXiv:gr-qc/0608122 [gr-qc]].

\bibitem{intro-Akbar:2007zz}
M.~Akbar and A.~A.~Siddiqui,
``Charged rotating BTZ black hole and thermodynamic behavior of field equations at its horizon,''
Phys. Lett. B \textbf{656}, 217-220 (2007)
doi:10.1016/j.physletb.2007.09.053
[arXiv:1009.3749 [gr-qc]].





\bibitem{M-Chen:2012mh}
B.~Chen, S.~x.~Liu and J.~j.~Zhang,
``Thermodynamics of Black Hole Horizons and Kerr/CFT Correspondence,''
JHEP \textbf{11}, 017 (2012)
doi:10.1007/JHEP11(2012)017
[arXiv:1206.2015 [hep-th]].

\bibitem{M-Cai:1996df}
R.~G.~Cai, Z.~J.~Lu and Y.~Z.~Zhang,
``Critical behavior in (2+1)-dimensional black holes,''
Phys. Rev. D \textbf{55}, 853-860 (1997)
doi:10.1103/PhysRevD.55.853
[arXiv:gr-qc/9702032 [gr-qc]].

\bibitem{M-Cai:1999dz}
R.~G.~Cai and J.~B.~Griffiths,
``Null particle solutions in three-dimensional (anti-)de Sitter spaces,''
J. Math. Phys. \textbf{40}, 3465-3475 (1999)
doi:10.1063/1.532900
[arXiv:gr-qc/9905011 [gr-qc]].

\bibitem{M-Cai:1998ep}
R.~G.~Cai and J.~H.~Cho,
``Thermodynamic curvature of the BTZ black hole,''
Phys. Rev. D \textbf{60}, 067502 (1999)
doi:10.1103/PhysRevD.60.067502
[arXiv:hep-th/9803261 [hep-th]].

\bibitem{M-Zou:2014gla}
D.~C.~Zou, Y.~Liu, B.~Wang and W.~Xu,
``Thermodynamics of rotating black holes with scalar hair in three dimensions,''
Phys. Rev. D \textbf{90}, no.10, 104035 (2014)
doi:10.1103/PhysRevD.90.104035
[arXiv:1408.2419 [hep-th]].


\bibitem{M-Wang:2006eb}
S.~Wang, S.~Q.~Wu, F.~Xie and L.~Dan,
``The First laws of thermodynamics of the (2+1)-dimensional BTZ black holes and Kerr-de Sitter spacetimes,''
Chin. Phys. Lett. \textbf{23}, 1096-1098 (2006)
doi:10.1088/0256-307X/23/5/009
[arXiv:hep-th/0601147 [hep-th]].

\bibitem{JT-Sadeghi:2015out}
J.~Sadeghi and A.~S.~Kubeka,
``P-V Criticality of Modified BTZ Black Hole,''
Int. J. Theor. Phys. \textbf{55}, no.5, 2455-2459 (2016)
doi:10.1007/s10773-015-2882-x


\bibitem{con-Liang:2021elg}
J.~Liang, W.~Lin and B.~Mu,
``Joule-Thomson expansion of the torus-like black hole,''
[arXiv:2103.03119 [gr-qc]].
\end{thebibliography}
\end{document}